\documentclass{article}

\usepackage{arxiv}
\usepackage[utf8]{inputenc}    
\usepackage[T1]{fontenc}        
\usepackage{hyperref}           
\usepackage{url}                    
\usepackage{booktabs}           
\usepackage{amsfonts}           
\usepackage{nicefrac}           
\usepackage{microtype}          
\usepackage{longtable}
\usepackage{graphicx}
\usepackage{subcaption}
\usepackage{algorithm}
\usepackage{capt-of}
\usepackage{xcolor}
\usepackage{mathtools}        
\usepackage{algpseudocode}
\usepackage[authoryear]{natbib}
\usepackage{bm}
\usepackage{enumitem}

\usepackage{pgfplots}
\usepackage{tikz}
\usetikzlibrary{calc,fadings,shapes.misc,3d,arrows,
decorations,decorations.pathmorphing,
decorations.text,shapes.geometric,intersections,
decorations.markings,patterns,spy,chains,positioning}
\usepgfplotslibrary{fillbetween,polar,patchplots}

\definecolor{myblue1}{RGB}{0,177,234}
\definecolor{myblue2}{RGB}{76,200,239}
\definecolor{myblue3}{RGB}{127,215,244}
\definecolor{myblue4}{RGB}{178,231,248}
\definecolor{mybluegray1}{RGB}{0,127,167}
\definecolor{mybluegray2}{RGB}{76,165,193}
\definecolor{mybluegray3}{RGB}{127,191,211}
\definecolor{mybluegray4}{RGB}{178,216,228}
\definecolor{mygray1}{RGB}{76,84,93}
\definecolor{mygray2}{RGB}{129,135,141}
\definecolor{mygray3}{RGB}{165,169,174}
\definecolor{mygray4}{RGB}{201,203,206}
\definecolor{mygray5}{RGB}{231,233,236}
\definecolor{myorange1}{RGB}{255,126,46}
\definecolor{myorange2}{RGB}{255,164,108}
\definecolor{myorange3}{RGB}{255,190,150}
\definecolor{myorange4}{RGB}{255,216,192}
\setenumerate{label=\textcolor{mybluegray1}{$\bm{\diamond}$},itemsep=1pt,topsep=5pt} 
\newcommand\ie{\textit{i.e.}\,}
\newcommand{\eps}{\varepsilon}
\title{A review on Deep Reinforcement Learning for Fluid Mechanics}

\author{
    Paul Garnier \\
     MINES Paristech , PSL - Research University \\
    \texttt{paul.garnier@mines-paristech.fr} \\
\And
     Jonathan Viquerat \\
    MINES Paristech , PSL - Research University\\
    CEMEF\\
    \texttt{jonathan.viquerat@mines-paristech.fr} \\
\And
    Jean Rabault\\
    Department of Mathematics, University of Oslo\\
        and\\
        MINES Paristech , PSL - Research University\\
        CEMEF\\
    \texttt{jean.rblt@gmail.com} \\
\And
    Aurélien Larcher\\
    MINES Paristech , PSL - Research University\\
    CEMEF\\
    \texttt{aurelien.larcherm@mines-paristech.fr} \\
\And
    Alexander Kuhnle\\
    University of Cambridge\\
    \texttt{alexkuhnle@t-online.de} \\
\And
    Elie Hachem\thanks{Corresponding author} \\
    MINES Paristech , PSL - Research University\\
    CEMEF\\
    \texttt{elie.hachem@mines-paristech.fr} \\
}

\begin{document}
\maketitle

\begin{abstract}
Deep reinforcement learning (DRL) has recently been adopted in a wide range of
physics and engineering domains for its ability to solve decision-making
problems that were previously out of reach due to a combination of
non-linearity and high dimensionality. In the last few years, it has
spread in the field of computational mechanics, and particularly in fluid
dynamics, with recent applications in flow control and shape optimization.
In this work, we conduct a detailed review of existing DRL applications to fluid
mechanics problems. In addition, we present recent results that
further illustrate the potential of DRL in Fluid Mechanics. The coupling
methods used in each case are covered, detailing their advantages and limitations. 
Our review also focuses
on the comparison with classical methods for optimal control
and optimization. Finally, several test cases are described that illustrate
recent progress made in this field. The goal of this publication is
to provide an understanding of DRL capabilities along with state-of-the-art
applications in fluid dynamics to researchers wishing to address new
problems with these methods.
\end{abstract}


\section{Introduction}

The process of learning and how the brain understands and classifies
information has always been a source of fascination for humankind. Since
the beginning of research in the field of artificial intelligence (AI), an
algorithm able to learn to make decisions on its own has been one of the
ultimate goals. Going back to the end of the 1980s \citep{Sutton1988},
reinforcement learning (RL) proposed a formal framework in which an agent
learns by interacting with an environment through the gathering of
experience \citep{Francois-lavet2018}. Significant results were obtained
during the following decade \citep{Tesauro1995}, although they were limited to
low-dimensional problems. 

In recent years, the rise of deep neural networks
(DNN) has provided reinforcement learning with new powerful tools
\cite{Goodfellow2017} \textcolor{black}{\cite{Strang2019} \cite{Bottou2018}}. The combination of deep learning with RL, called deep
reinforcement learning (DRL), lifted several major obstacles that hindered
classical RL by allowing the use of high-dimensional state spaces and
exploiting the feature extraction capabilities of DNNs. Recently, DRL managed
to perform tasks with unprecedented efficiency in many domains such as robotics
\cite{Pinto2017} or language processing \cite{Bahdanau2016}, and even achieved
superhuman levels in multiple games: Atari games \cite{Mnih2013}, Go
\cite{Silver2017}, Dota II \cite{OpenAI2018}, Starcraft II \cite{Vinyals2019}
and even Poker \cite{Browneaay2400}. Also in industrial application, DRL
has proven itself useful. For example, \textit{Wayve}
trains autonomous cars both onboard \citep{Kendall2018} and through simulation
\citep{Bewley2018}, and Google uses DRL in order to
control the cooling of its data centers \citep{GoogleDataCenter2018}.

In the field of fluid mechanics and mechanical engineering, one is also
confronted with nonlinear problems of high dimensionality. For example, using
computational simulations to test several different designs or configurations
has proven a useful technique. However, the number of possibilities to explore can make
such search difficult since it is often unfeasible to evaluate all configurations
exhaustively. Therefore, the assistance of automatic optimization procedures is
needed to help find optimal designs.

Possible applications range from multiphase flows in micro-fluidics to shape
optimization in aerodynamics or conjugate heat transfer in heat exchangers, and
many other fields can gain from such techniques such as biomechanics, energy, or marine technologies. This is the
primary motivation for the combination of computational fluid dynamics (CFD)
with numerical optimization methods. Most of the time, 
the main obstacle is the high computational cost of determining the
sensitivity of the objective function to variations of the design parameters by
repeated calculations of the flow. Besides, classical optimization techniques,
like direct gradient-based methods,
are known for their lack of robustness and for
their tendency to fall into local optima. Generic and robust search methods,
such as evolutionary or genetic algorithms, offer several
attractive features and have been used widely for design shape optimization
\citep{Ali2003, Lee1996, Makinen1999}. In particular, they can
be used for multi-objective multi-parameter problems. They have been
successfully tested for many practical cases, for example for design shape
optimization in the aerospace \citep{Matos2004} and automotive industries
\citep{Muyl2004}. However, the use of a fully automatic evolutionary algorithms coupled with CFD for the optimization
of multi-objective problems remains limited by the computing burden at stake and is still
far from providing a practical tool for many engineering applications.

Despite great potential, the literature applying DRL to computational or experimental
fluid dynamics remains limited. In recent years, only a handful of applications
were proposed that take advantage of such learning methods. These applications
include flow control problems \citep{Rabault2019, Gueniat2016}, process
optimization \citep{Novati2017, Lee2018}, shape optimization 
\citep{Lampton2018}, and our own
results about laminar flows past a square cylinder presented later in this work, among
others\footnote{These references and more are addressed in details in the
remaining of this article.}.

This paper presents an introduction to modern RL for the neophyte
reader to get a grasp of the main concepts and methods existing in the domain,
including the necessary basics of deep learning. Subsequently, the relevant
fluid mechanics concepts are reiterated. We then go through a 
thorough review of the literature and describe the content of each contribution.
In particular, we cover the choice of DRL algorithm, the problem
complexity, and the concept of reward shaping. Finally, an outline is given on
the future possibilities of DRL coupled with fluid dynamics.

\section{Deep Reinforcement Learning}

In this section, the basic concepts of DRL are introduced. For the sake of
brevity, some topics are only minimally covered, and the reader is referred
to more detailed introductions to (D)RL \citep{Sutton2018, Francois-lavet2018}.
First, RL is presented in its mathematical description as a Markov
Decision Process. Then, its combination with deep learning is
detailed, and an overview of several DRL algorithms is presented.

\begin{table}
    \footnotesize
    \caption{\textbf{Notations regarding DRL}}
    \label{tab:drl}
    \centering
    \begin{tabular}{cl}
        \toprule
        $\gamma$ & discount factor\\
        $\lambda$ & learning rate\\
        $\alpha$, $\beta$ & step-size\\
        $\epsilon$ & probability of random action\\\midrule
        $s$,$s'$ & states\\
        $\mathcal{S}^{+}$ & set of all states\\
        $\mathcal{S}$ & set of non-termination states\\
        $a$ & action\\
        $\mathcal{A}$ & set of all actions\\
        $r$ & reward\\
        $\mathcal{R}$ & set of all rewards\\
        $t$ & time station\\
        $T$ & final time station\\
        $a_t$ & action at time $t$\\
        $s_t$ & state at time $t$\\
        $r_t$ & reward at time $t$\\
        $R(\tau)$ & discounted cumulative reward following trajectory $\tau$\\
        $R(s,a)$ & reward received for taking action $a$ in state $s$\\\midrule
        $\pi$ & policy\\
        $\theta$, $\theta'$ & parameterization vector of a policy\\
        $\pi_\theta$ & policy parameterized by $\theta$\\
        $\pi (s)$ & action probability distribution in state $s$ following $\pi$\\
        $\pi (a \vert s)$ & probability of taking action $a$ in state $s$ following $\pi$\\
        $V^{\pi}(s)$ & value of state $s$ under policy $\pi$\\
        $V^{*}(s)$ & value of state $s$ under the optimal policy\\
        $Q^{\pi}(s,a)$ & value of taking action $a$ in state $s$ under policy $\pi$\\
        $Q^{*}(s,a)$ & value of taking action $a$ in state $s$ under the optimal policy\\
        $Q_\theta(s,a)$ & estimated value of taking action $a$ in state $s$ with parameterization $\theta$\\\bottomrule
    \end{tabular}
\end{table}

\subsection{Reinforcement Learning}

\begin{figure}
    \centering
    \includegraphics[width=.6\linewidth]{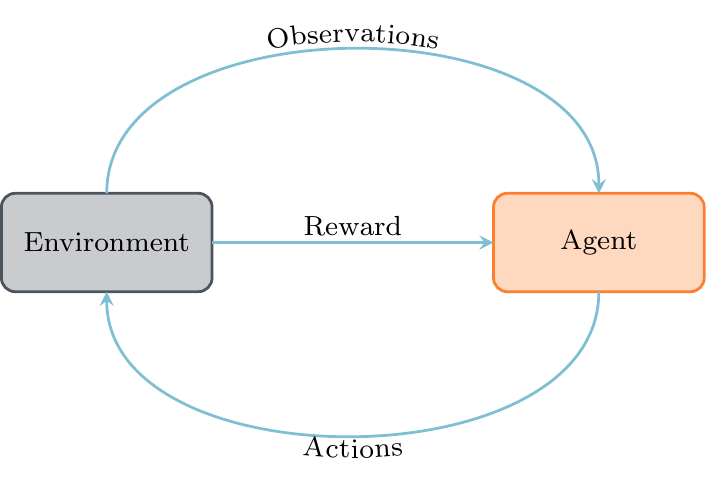}
    \caption{\textbf{DRL agent and its environment}}
    \label{fig:summary_rl}
\end{figure}

At its core, reinforcement learning models an agent
interacting with an environment and receiving observations and rewards from it, 
as shown in figure \ref{fig:summary_rl}.
The goal of the agent is to determine the best action in any given state in order to maximize
its cumulative reward over an episode, \ie over one instance of the scenario in which the agent takes actions. 
Taking the illustrative example of a board game: 

\begin{enumerate}
	\item An episode is equivalent to one game;
	\item A state consists of the summary of all pieces and their positions;
	\item Possible actions correspond to moving a piece of the game;
	\item The reward may simply be $+1$ if the agent wins, $-1$ if it loses and $0$
		in case of a tie or a non-terminal game state.
\end{enumerate}

Generally, the reward is a signal of how \emph{`good'} a board situation is, 
which helps the agent to learn distinguish more promising from less attractive decision trajectories.
A trajectory is a sequence of states and actions experienced by the agent:

\begin{equation*}
	\tau = \big( s_0, a_0, s_1, a_1, ... \big).
\end{equation*}

The cumulative reward (\ie the quantity to maximize) is expressed along a trajectory, and 
includes a discount factor $\gamma \in \left[0,1\right]$ that 
smoothes the impact of temporally distant rewards (it is then called \emph{discounted} cumulative 
reward):

\begin{equation*}
	R(\tau) = \displaystyle \sum_{t=0}^{T} \gamma^{t} r_{t}.
\end{equation*}

RL methods are classically divided in two categories, namely \emph{model-free}
and \emph{model-based} algorithms. Model-based method incorporates a model of
the environment they interact with, and will not be considered in this paper \textcolor{black}{(the reader is referred to \cite{Sutton2018} and references therein for details about model-based methods)}.
On the contrary, model-free algorithms directly interact with their
environment, and are currently the most commonly used within the DRL community,
mainly for their ease of application and implementation. Model-free methods are
further distinguished between \emph{value-based} methods and
\emph{policy-based} methods \citep{Sutton1998}.  Although both approaches aim
at maximizing their expected return, policy-based methods do so by directly
optimizing the decision policy, while value-based methods learn to estimate the
expected value of a state-action pair optimally, which in turn determines the
best action to take in each state.  Both families inherit from the formulation
of Markov decision processes (MDP), which is detailed in the following section.

\subsubsection{Markov decision processes}

A MDP can be defined by a tuple $(\mathcal{S},\mathcal{A},\mathbb{P}, R)$ \citep{Bellman1957,Howard1960}, where:

\begin{enumerate}
    \item $\mathcal{S}$ , $\mathcal{A}$ and $R$ are defined in table \ref{tab:drl},
    \item $\mathbb{P} \colon \mathcal{S} \times \mathcal{A} \times \mathcal{S}^{+} \to [0,1]$, 
    where $\mathbb{P} (s' \vert s, a)$ is the probability of getting to state $s'$ from state $s$ following action $a$.
\end{enumerate}

In the formalism of MDP, the goal is to find a policy $\pi(s)$ that maximizes the expected reward. 
However, in the context of RL, $\mathbb{P}$ and $R$ are unknown to the agent, which is expected 
to come up with an efficient decisional process by interacting with the environment. The way the 
agent induces this decisional process classifies it either in value-based or policy-based methods.

\subsubsection{Value-based methods}

In value-based methods, the agent learns to optimally estimate a \emph{value function}, 
which in turn dictates the policy of the agent by selecting the action of the highest value. 
One usually defines the \emph{state value function}:

\begin{equation*}
    V^\pi (s) = \underset{\tau \sim \pi}{\mathbb{E}} \big[ R(\tau) \vert s \big],
\end{equation*}

\noindent and the \emph{state-action value function}:

\begin{equation*}
    Q^\pi (s,a) = \underset{\tau \sim \pi}{\mathbb{E}} \big[ R(\tau) \vert s, a \big],
\end{equation*}

\noindent which respectively denote the expected discounted cumulative reward starting in state 
$s$ (resp. starting in state $s$ and taking action $a$) and then follow trajectory $\tau$ 
according to policy $\pi$. It is fairly straightforward to see that these two concepts are 
linked as follows:

\begin{equation*}
    V^\pi (s) = \underset{a \sim \pi}{\mathbb{E}} \big[ Q^\pi (s, a) \big],
\end{equation*}

\noindent meaning that in practice, $V^\pi (s)$ is the weighted average of $Q^\pi (s, a)$ over all 
possible actions by the probability of each action. Finally, the 
\emph{state-action advantage function} can be defined as $A^{\pi} (s,a) = Q^\pi (s,a) - V^\pi (s)$. 
One of the main value-based 
methods in use is called Q-learning, as it relies on the learning of the Q-function to 
find an optimal policy. In classical Q-learning, the Q-function is stored in a Q-table, 
which is a simple array representing the estimated value of the optimal Q-function 
$Q^*(s, a)$ for each pair $(s,a) \in \mathcal{S} \times \mathcal{A}$. The Q-table is 
initialized randomly, and its values are progressively updated as the agent explores 
the environment, until the Bellman optimality condition \citep{Bellman1962} is reached: 

\begin{equation}
\label{eq:bellman}
    Q^* (s,a) = R(s,a) + \gamma \max_{a'} Q^* (s',a').
\end{equation}

The Bellman equation indicates that the Q-table estimate of the Q-value has 
converged and that systematically taking the action with the highest Q-value leads 
to the optimal policy. In practice, the expression of the Bellman equation \citep{Bellman1962} is used 
to update the Q-table estimates.

\subsubsection{Policy-based methods}
\label{section:policy}

Another approach consists of directly optimizing a parameterized policy 
$\pi_\theta(a \vert s)$, instead of resorting to a value function estimate 
to determine the optimal policy. In contrast with value-based methods, 
policy-based methods offer three main advantages:

\begin{enumerate}
	\item They have better convergence properties, although they tend to be 
	trapped in local minima;
	\item They naturally handle high dimensional action spaces;
	\item They can learn stochastic policies.
\end{enumerate}

To determine how "good" a policy is, one defines an objective function based on 
the expected cumulative reward:

\begin{equation*}
    J(\theta) = \underset{\tau \sim \pi_\theta}{\mathbb{E}} \big[ R(\tau) \big],
\end{equation*}

\noindent and seeks the optimal parameterization $\theta^*$ that maximizes $J(\theta)$:

\begin{equation*}
    \theta^* = \arg \max_\theta \mathbb{E} \big[ R(\tau) \big].
\end{equation*}

To that end, the gradient of the cost function is needed. This is not an obvious
task at first, as one is looking for the gradient with respect to the policy parameters 
$\theta$, in a context where the effects of policy changes on the state distribution 
are unknown. Indeed, modifying the policy will most certainly modify the set of
visited states, which could affect performance in an unknown manner. 
This derivation is a classical exercise that relies on
the log-probability trick \textcolor{black}{\cite{Williams1992}}, which allows expressing $\nabla_\theta J(\theta)$ as an 
evaluable expected value:

\begin{equation}
\label{eq:policy_gradient}
	\nabla_\theta J(\theta) = \underset{\tau \sim \pi_\theta}{\mathbb{E}} \left[ \sum_{t=0}^T \nabla_\theta \log \left( \pi_\theta (a_t \vert s_t) \right) R(\tau) \right].
\end{equation}

This gradient is then used to update the policy parameters:

\begin{equation}
\label{eq:theta_update}
	\theta \leftarrow \theta + \lambda \nabla_\theta J(\theta).
\end{equation}

As $\nabla_\theta J(\theta)$ takes the form of expected value, in practice, 
it is evaluated by averaging its argument over a set of trajectories. 
In such vanilla policy gradient methods, the actions are evaluated at the end 
of the episode (they belong to Monte-Carlo methods). If some low-quality 
actions are taken along the trajectory; their negative impact will be averaged 
by the high-quality actions and will remain undetected. This problem is 
overcome by actor-critic methods, in which a Q-function evaluation is used 
in conjunction with a policy optimization.

\subsection{Deep Reinforcement Learning}

In RL, the evaluation of value functions such as $V^\pi$ or $Q^\pi$ can be done 
in several ways. One possible way, as presented earlier, is to use tables that 
store the values for every state or action-state pair. However, this strategy does 
not scale with the sizes of state and action spaces. Another possibility is to use 
neural networks to estimate value functions, or, for policy-based methods, to output 
an action distribution given input states. Artificial neural networks can be seen 
as universal function approximators \citet{Siegelmann1995132}, which means 
that, if properly trained, they can represent arbitrarily complex mappings between
spaces. This particularly suitable property of ANNs gave rise to deep reinforcement
learning, \ie reinforcement learning algorithms using deep neural networks as
functions approximators. Regarding the notions introduced in previous sections, 
it is natural to think of $\theta$ as the parameterization of the neural network, \ie 
as it set of weights and biases. In the following section, a basic introduction to 
neural networks is given. The reader is referred to \cite{Goodfellow2017} for 
additional details.

\subsubsection{Artificial neurons and neural networks}
\label{section:ANN}

\begin{figure}
\newsavebox{\largestimage}
	\centering
	\savebox{\largestimage}{\includegraphics[width=.45\textwidth]{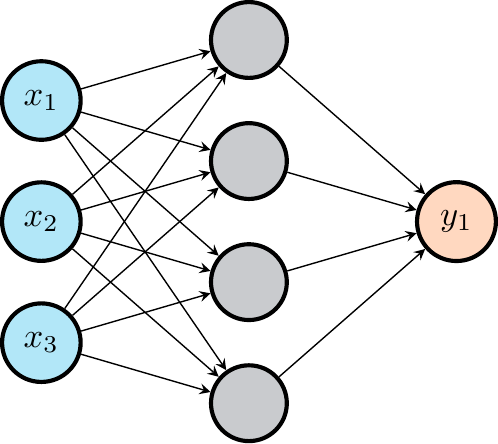}}
	\begin{subfigure}[t]{.45\textwidth}
		\centering
		\raisebox{\dimexpr.55\ht\largestimage-.5\height}{%
			\includegraphics[width=\textwidth]{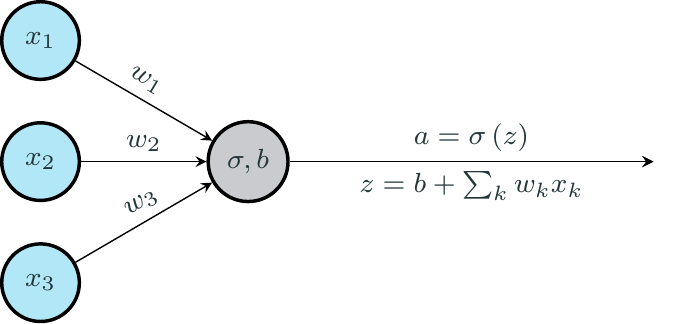}
		}
		\caption{	Representation of a single artificial neuron. The neuron receives
    				an input vector $x$, assorted with a weight vector $w$. The output is
				computed as $w \cdot x$ corrected with the bias $b$, to which is applied
				the activation function $\sigma$.}
	\end{subfigure}
	\qquad
	\begin{subfigure}[t]{.45\textwidth}
		\centering
		\usebox{\largestimage}
		\caption{	Simple example of neural network with an input vector $x \in
				\mathbb{R}^3$, a hidden layer composed of $4$ neurons, and an output layer
				composed of a single neuron. As a convention, input variables are drawn
				using a neuron representation. However, it must be kept in mind that the
				input layer is not composed of neurons.}
	\end{subfigure}
	\caption{\textbf{Representation of a single artificial neuron and a basic 3-layer neural network}.}
	\label{fig:NN_schema}
\end{figure}

The basic unit of a neural network (NN) is the neuron, which representation is given in figure
\ref{fig:NN_schema}. An input vector $x$, associated with a set of weights
$w$, is provided to the neuron. The neuron then computes the weighted sum $w
\cdot x + b$, where $b$ is called the bias and applies the activation function
$\sigma$ to this sum. This is the output of the neuron, hereafter noted $z$. In
the neuron, the weights and the bias represent the degrees of freedom (\ie,
the parameters that can be adjusted to approximate the function $f$), while the
activation function is a hyper-parameter, \ie, it is part of the choices made
during the network design.

In their purest form, neural networks consist of several layers of neurons
connected to each other, as shown in the primary example of \ref{fig:NN_schema}. Such a
network is said to be fully connected (FC), in the sense that each neuron of a
layer is connected to all the neurons of the following layer. As it was said in
the last section, each connection is characterized by a weight, and in addition each neuron has a bias, except
for the input layer. Indeed, as a convention, the input layer is usually drawn
as a regular layer, but it does not hold biases, nor is an activation function
applied at its output. The learning process in neural networks consists in 
adjusting all the biases and weights of the network in order to reduce the 
value of a well-chosen loss function that represents the quality of the network 
prediction. This update is usually performed by a stochastic gradient method, 
in which the gradients of the loss function with respect to the weights and biases 
are estimated using a back-propagation algorithm.

When working for instance with images as input, it is customary to exploit
convolutional layers instead of FC ones. Rather than looking for patterns in
their entire input space as FC layers, convolutional ones can extract local
features. Additionally, they can build a hierarchy of increasingly
sophisticated features. In such layers, a convolution kernel (\ie, a tensor
product with a weight matrix) is applied on a small patch of the input image
and is used as input for a neuron of the next layer. The patch is then moved,
and the operation repeated until the input image has been entirely covered. The
size of the patch is usually known as the kernel size. The complete coverage of
the image with this process generates a kernel, also called filter or feature
map. In most cases, multiple kernels are generated at each layer, each encoding
a specific feature of the input image. When used for regression applications,
convolutional networks (or \textit{convnets}) most often end with a fully connected
layer, followed by the output layer, which size is determined by that of the
sought quantity of interest. These considerations are discussed in details in a
large variety of books and articles such as \cite{Goodfellow2017} \cite{Strang2019} \cite{Bottou2018}.

\subsection{DRL algorithms}

In this section, the main lines of the major DRL algorithms in use today are 
presented. For the sake of brevity, many details and discussions are absent from 
the following discussions. The reader is referred to the profuse literature on the topic.

\subsubsection{Deep Q-networks}
\label{section:DQN}

Instead of updating a table holding Q-values for each possible $(s,a)$ pair, 
ANNs can be used to generate a 
map $\mathcal{S}^{+} \times \mathcal{A} \longrightarrow \mathbb{R}$ (called 
deep Q-networks (DQN)) that provides an estimate of the Q-value for each possible action 
given an input state. As it is usual for neural networks, the update of the 
Q-network requires a loss function for the gradient descent algorithm, 
that will be used to optimize the network parameters $\theta$. To do so, 
it is usual to exploit the Bellman optimality equation (\ref{eq:bellman}):

\begin{equation}
\label{eq:DQN_loss}
	L(\theta) = \mathbb{E} \Bigg[ \frac{1}{2} \Big(\underbrace{ \big[ R(s,a) + \gamma \max_{a'} Q_\theta (s',a') \big]}_\text{Target} - \,Q_\theta(s,a) \Big)^2 \Bigg].
\end{equation}

\noindent In the latter expression, $Q_\theta(s,a)$ represents the Q-value \emph{estimate} 
provided by the DQN for action $s$ and state $a$ under network parameterization $\theta$.
It must be noted that the quantity denoted \emph{target} is the same that appears
in the Bellman optimality equation (\ref{eq:bellman}): when the optimal set of 
parameters $\theta^*$ is reached, $Q_\theta(s, a)$ is equal to the target, and $L(\theta)$ 
is equal to zero.

\noindent In vanilla deep Q-learning (and more generally in Q-learning), an 
exploration/exploitation trade-off must be implemented, to ensure a sufficient 
exploration of the environment. To do so, a parameter $\epsilon \in [0,1]$ is defined,
and a random value is drawn in the same range before each action. If this value
is below $\epsilon$, a random action is taken. Otherwise, the algorithm follows the
action prescribed by $\max_a Q_\theta(s, a)$. Most often, $\epsilon$ follows a schedule,
starting with high values at the beginning of learning, and progressively decreasing.
The vanilla algorithm using DQN is shown below : 

\begin{algorithm}
	\caption{Vanilla deep Q-learning}
	\begin{algorithmic}[1]
		\State Initialize action-value function parameters $\theta_0$
		\For {$k$ = $0$, $M$} \Comment{Loop over episodes}
			\For {$t$ = $0$, $T$} \Comment{Loop over time stations}
				\State Draw random value $\omega \in [0,1]$ \Comment{$\epsilon$-greedy strategy}
				\If {($\omega < \epsilon$)}
					\State Choose action $a_t$ randomly
				\Else
					\State Choose action $a_t = \max_a Q_\theta(s_t, a)$
				\EndIf
				\State Execute action $a_t$ and observe reward $r_t$ and state $s_{t+1}$
				\State Form target $y_t = r_t + \max_a Q_\theta(s_{t+1}, a)$
				\State Perform gradient descent using $(y_t - Q_\theta(s_t, a_t))^2$
				\State Update $\theta$ and $s_t$
			\EndFor 
		\EndFor
	\end{algorithmic}
\end{algorithm}

Still, the basic deep Q-learning algorithm presents several flaws, and many improvements 
can be added to it:

\begin{enumerate}
	\item \textbf{Experience replay}: To avoid forgetting about its past learning, 
	a replay buffer can be created that contains random previous experiences.
	This buffer is regularly fed to the network as learning material, so previously 
	acquired behaviors are not erased by new ones. This improvement also reduces
	the correlation between experiences: as the replay buffer is randomly shuffled, 
	the network experiences past $(s_t, a_t, r_t, s_{t+1})$ tuples in a different order,
	and is therefore less prone to learn correlation between these 
	\citep{Tsitsiklis1997,Lin1993};
	\item \textbf{Fixed Q-targets}: When computing the loss (\ref{eq:DQN_loss}), the 
	same network is used to evaluate the target and $Q_\theta(s_t, a_t)$. Hence, at 
	every learning step, both the Q-value and the target move, \ie the gradient descent 
	algorithm is chasing a moving target, implying big oscillations in the training. To 
	overcome this, a separate network is used to evaluate the target (target network), 
	and its weight parameters $\theta^-$ are updated less frequently than that of the 
	Q-network. This way, the target remains fixed for several Q-network updates, making 
	the learning easier;
	\item \textbf{Double DQN}: The risk of using a single Q-network is that it may 
	over-estimate the Q-values, leading to sub-optimal policies, as $\pi_\theta$ is 
	derived by systematically choosing the actions of highest Q-values. In \citep{DDQN},
	two networks are used jointly: a DQN is used to select the best action, while the 
	other one is used to estimate the target value;
	\item \textbf{Prioritized replay buffer}: In the exploration of the environment by the 
	agent, some experiences might occur rarely while being of high importance. As the 
	memory replay batch is sampled uniformly, these experiences may never be selected. 
	In \cite{PrioritizedER}, the authors propose to sample the replay buffer according to 
	the training loss value, in order to fill the buffer in priority with experiences where the 
	Q-value evaluation was poor, meaning that the network will have better training on 
	"unexpected events". To avoid overfitting on these events, the filling of the prioritized 
	buffer follows a stochastic rule.
\end{enumerate}

\subsubsection{Deep policy gradient}
\label{section:DPG}

When the policy $\pi_\theta$ is represented by a deep neural network with parameters 
$\theta$, the evaluation of $\nabla_\theta J(\theta)$ can be delegated to the 
back-propagation algorithm (see section \ref{section:ANN}), as long as the gradient of 
the loss function equals the policy gradient (\ref{eq:policy_gradient}). The loss can be 
expressed as:

\begin{equation*}
	L(\theta) = \underset{\tau \sim \pi_\theta}{\mathbb{E}} \left[ \sum_{t=0}^T \log \left( \pi_\theta (a_t \vert s_t) \right) R(\tau) \right].
\end{equation*}

\noindent However, it can be shown that using the full history of discounted reward along trajectory 
$\tau$ is not necessary, as a given action $a_t$ has no influence on the rewards obtained 
at previous time-stations. Several expressions are possible to use \textcolor{black}{instead} of $R(\tau)$, 
that do not modify the value of the computed policy gradient, while reducing its variance (and 
therefore the required amount of trajectories to correctly approximate the expected value) 
\citep{Schulman2015}. Among those, the advantage function $A(s,a)$ is commonly chosen for 
its low variance:

\begin{equation*}
	A (s,a) = Q (s,a) - V (s).
\end{equation*}

\noindent The advantage function $A(s,a)$, represents the improvement obtained 
in the expected cumulative reward when taking action $a$ in state $s$, compared to 
the average of all possible actions taken in state $s$. In practice, it is not readily available 
and must be estimated using the value function. 
In practice, this estimate is provided by a separately learned value function.
As a result, the loss function writes as:

\begin{equation*}
	L(\theta) = \underset{\tau \sim \pi_\theta}{\mathbb{E}} \left[ \sum_{t=0}^T \log \left( \pi_\theta (a_t \vert s_t) \right) A^{\pi_\theta} (s_t, a_t) \right].
\end{equation*}

\noindent The vanilla deep policy gradient algorithm is presented below:

\begin{algorithm}
	\caption{Vanilla deep policy gradient}
	\begin{algorithmic}[1]
		\State Initialize policy parameters $\theta_0$
		\For {$k$ = $0$, $M$} \Comment{Loop over batches of trajectories}
			\For {$i$ = $0$, $m$} \Comment{Loop over trajectories in batch $k$}
				\State Execute policy $\pi_{\theta_k}$
				\State Collect advantage estimates $A^{\pi_{\theta_k}} (s_t, a_t)$
			\EndFor 
			\State Estimate $L(\theta_k)$ over current batch of trajectories
			\State Update policy parameters $\theta_k$ using stochastic gradient ascent
		\EndFor
	\end{algorithmic}
\end{algorithm}

\subsubsection{Advantage actor-critic}

As mentioned in section \ref{section:policy}, actor-critic methods propose to 
simultaneously exploit two networks to improve the learning performance. One of 
them is policy-based, while the other one is value-based:

\begin{enumerate}
	\item The actor $\pi_\theta (s,a)$, controls the actions taken by the agent;
	\item The critic $Q_w(s,a)$, evaluates the quality of the action taken by the actor.
\end{enumerate}

\noindent Both these networks are updated in parallel, in a time-difference (TD) fashion (one 
update at each time station) instead of the usual Monte-Carlo configuration (one 
update at the end of the episode). In practice, the advantage function $A(s,a)$ is preferred 
to the Q-value for its lower variability. In order to avoid having to evaluate two value functions, 
the advantage function is usually 
approximated, using the reward obtained by the actor after action $a_t$ and the value evaluated 
by the critic in state $s_{t+1}$:

\begin{equation*}
	A_w(s_t,a_t) \sim R(s_t,a_t) + \gamma V_w(s_{t+1}) - V_w(s_t).
\end{equation*}

\noindent The actor-critic process then goes as follows:

\begin{algorithm}
	\caption{Advantage actor-critic}
	\begin{algorithmic}[1]
		\State Initialize policy parameters $\theta_0$ \Comment{Actor}
		\State Initialize action-value function parameters $w_0$ \Comment{Critic}
		\For {$k$ = $0$, $M$} \Comment{Loop over episodes}
			\For {$t$ = $0$, $T$} \Comment{Loop over time stations}
				\State Execute action $a_t$ following $\pi_\theta$, get reward $r_t$ and state $s_{t+1}$ \Comment{Actor}
				\State Evaluate target $R(s_t,a_t) + \gamma V_w(s_{t+1})$  \Comment{Critic}
				\State Update $w$ to $w'$ using target \Comment{Critic}
				\State Evaluate advantage $A_{w'}(s_t, a_t) = R(s_t,a_t) + \gamma V_{w'}(s_{t+1}) - V_{w'}(s_t)$  \Comment{Critic}
				\State Update $\theta$ to $\theta'$ using $A_{w'}(s_t, a_t)$ \Comment{Actor}
			\EndFor 
		\EndFor
	\end{algorithmic}
\end{algorithm}

In the latter algorithm, the actor and critic parameter updates follow the methods described 
in sections \ref{section:DQN} and \ref{section:DPG}. 
The advantage actor critic algorithm has been developed in asynchronous (A3C \cite{Mnih2016}) 
and synchronous (A2C) versions. The version of the algorithm presented here corresponds to a 
time-difference A2C, although many possible variants exist, such as the n-step A2C.

\subsubsection{Trust-region and proximal policy optimization}

Trust region policy optimization (TRPO) was introduced in 2015 as a major improvement of vanilla 
policy gradient \cite{Schulman2015}. In this method, the network update exploits a 
\emph{surrogate advantage} functional:

\begin{equation*}
	\theta_{k+1} = \arg \max_{\theta} L(\theta_k, \theta),
\end{equation*}

\noindent with

\begin{equation*}
	L(\theta_k, \theta) = \underset{(s,a) \sim \pi_{\theta_k}}{\mathbb{E}} \big[ \Pi(s,a,\theta,\theta_k) A^{\pi_{\theta_k}} (s,a) \big],
\end{equation*}

\noindent and

\begin{equation*}
	\Pi(s, a, \theta, \theta_k) = \frac{\pi_\theta (a \vert s)}{\pi_{\theta_k} (a \vert s)}.
\end{equation*}
	
\noindent In latter expression, $L(\theta_k, \theta)$ measures how much better (or worse) 
the policy $\pi_{\theta}$ performs compared to the previous policy $\pi_{\theta_k}$. In order to avoid too
large policy updates that could collapse the policy performance, TRPO leverages second-order 
natural gradient optimization to update parameters within a trust-region of a fixed maximum 
Kullback-Leibler divergence between old and updated policy distribution. This relatively
complex approach was replaced in the PPO method by simply clipping the maximized
expression:

\begin{equation*}
	L(\theta_k, \theta) = \underset{(s,a) \sim \pi_{\theta_k}}{\mathbb{E}} \big[ \min \left( \Pi(s,a,\theta,\theta_k) A^{\pi_{\theta_k}} (s,a), g\left( \eps, A^{\pi_{\theta_k}} (s,a) \right) \right) \big],
\end{equation*}

\noindent where

\begin{equation*}
	g(\eps, A) = 
	\begin{cases}
		(1+\eps) A 	& A \geq 0,\\
		(1-\eps) A 		& A < 0,
	\end{cases}
\end{equation*}

\noindent where $\eps$ is a small, user-defined parameter. When $A^{\pi_{\theta_k}} (s,a)$ is positive, 
taking action $a$ in state $s$ is preferable to the average of all actions that could be taken in that state, 
and it is natural to update the policy to favor this action. Still, if the ratio $\Pi(s,a,\theta,\theta_k)$ is very large, 
stepping too far from the previous policy $\pi_{\theta_k}$ could damage performance. For that reason, 
$\Pi(s,a,\theta,\theta_k)$ is clipped to $1+\eps$ to avoid too large updates of the policy. If 
$A^{\pi_{\theta_k}} (s,a)$ is negative, taking action $a$ in state $s$ represents a poorer choice than 
the average of all actions that could be taken in that state, and it is natural to update the policy to 
decrease the probability of taking this action. In the same fashion, $\Pi(s,a,\theta,\theta_k)$ is clipped 
to $1-\eps$ if it happens to be lower than that value. 

Many refinements can be added to these methods that are out of reach of this introduction, 
such as the use of generalized advantage estimator \citep{Schulman2016}, the introduction 
of evolution strategies in the learning process \citep{Houthooft2018}, or the combination of 
DQN with PG methods for continuous action spaces \citep{Lillicrap2015}.

\subsection{A conclusion on DRL}

\begin{figure}
	\centering
	\includegraphics[width=.6\linewidth]{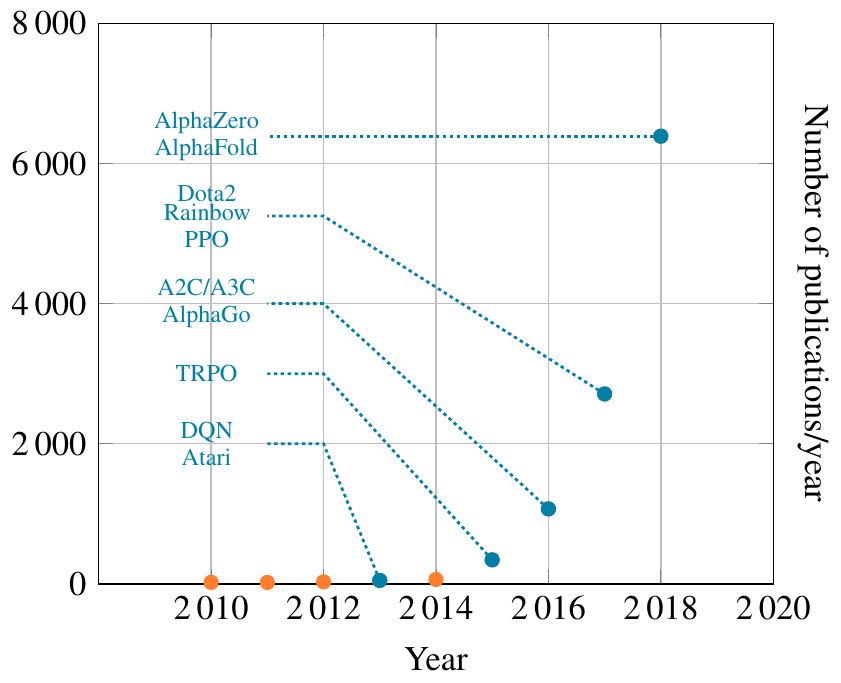}
	\caption{\textbf{Number of publications mentioning "Deep reinforcement learning" per year} since 2010 (data from \url{www.scholar.google.com}). In blue, we indicate some of the most important algorithms or achievements of the field.}
	\label{fig:nb_DRL_pub}
\end{figure}

Since the first edition of the book of Sutton \cite{Sutton1998}, RL has become a 
particularly active field of research. However, the domain really started striving 
after the paper from \cite{Mnih2013} and its groundbreaking results using DQN to 
play Atari games. As of today, the domain holds countless major achievements, 
and the ability of DRL to learn to achieve complex tasks is well established.
It is now exploited in multiple fields of research, and is featured in an ever-increasing amount 
of publications, as shown in figure \ref{fig:nb_DRL_pub}.

A major advantage of DRL is that it can be used as an agnostic tool for 
control and optimization tasks, both in continuous and discrete contexts. 
Its only requirement is a well-defined $(s_t, a_t, r_t)$ interface from the 
environment, which can consist in a numerical simulation or a real-life experiment.
As seen earlier, both DQN and DPG ensure a good exploration 
of the states space, making them able to unveil new optimal behaviors. 
Moreover, DRL is also robust to transfer learning, meaning that an 
agent trained on one problem will train much faster on a similar 
environment. This represents a great strength, especially for domains like 
fluid mechanics, where computational time represents a limiting factor (section 
\ref{section:opt_CC} illustrates this point). Transfer learning also introduces a 
large contrast between DRL and regular optimization techniques, such as adjoint methods, 
where the product of optimization in a given case cannot be re-used to speed up 
optimization in a similar configuration. Finally, as illustrated in section \ref{section:DRL_parallel}, 
DRL libraries can exploit parallel learning with a close-to-perfect scaling on the 
available resources. Hence, for its ease of use and its numerous capabilities, 
DRL represents a promising tool for optimization and 
design processes involving computational or experimental environments.

\newpage

\section{Applications}

In this section, several applications combining DRL and fluid mechanics found in the 
literature are presented in details. For each case, the numerical experiments and the 
obtained results are detailed. The choice of DRL algorithm and the problem complexity 
are also considered. Table \ref{tab:articles} presents an overview of the reviewed articles 
and their corresponding references.

\begin{table}[h!]
	\footnotesize
	\caption{\textbf{Classification of the papers scanned in this review}. ADQN and DDPG respectively stand for asynchronous DQN and deep deterministic PG.}
	\medskip
	\label{tab:articles}
	\centering
	\begin{tabular}{lll}
																						\toprule
		\textbf{Main topic}             		& \textbf{Algorithm}	& \textbf{References}						\\\midrule
		Flow Control                 		& QL                 		& \cite{Gueniat2016} 						\\
                 		              			& DQN              	& \cite{Gazzola2016}, \cite{Novati2017} 			\\
                               					& ADQN     		& \cite{Verma2018} 							\\
                               					& TRPO            	& \cite{Ma2018} 							\\
               			                			& A3C               		& \cite{garnier2019-controlcylinder} 				\\
                               					& PPO              		& \cite{Rabault2019}, \cite{RabaultKuhnle2019} 	\\\midrule
		Homogeneity optimization    	& QL                 		& \cite{Colabrese2017}, \cite{Gustavsson2017} 	\\
                            					& QL 			&  \cite{Qiu2018}, \cite{Tsang2018} 				\\\midrule
		Shape Optimization             	& QL  			& \cite{Lampton2018} 						\\\midrule
		Chaotic Systems      			& DDPG           		& \cite{Bucci2019} 							\\\bottomrule 
	\end{tabular}
\end{table}

\subsection{Synchronised swimming of two fish - \citet{Novati2017}}

\begin{figure}
	\centering
	\fbox{\includegraphics[width=\linewidth]{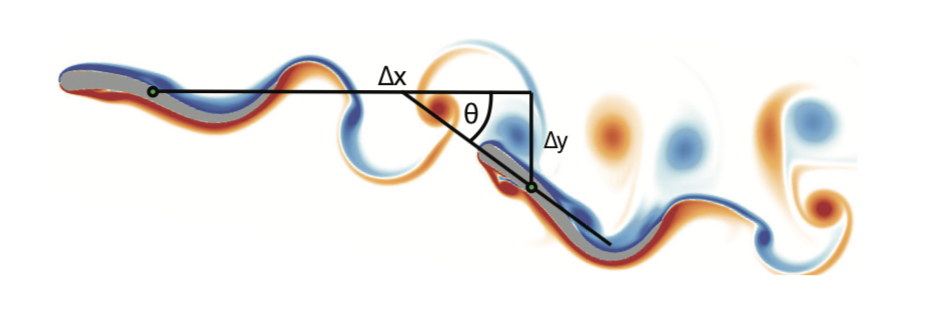}}
	\caption{\textbf{Leader and follower swimmer, reproduced from \cite{Novati2017}}, 
				with the displacements $\Delta x$ and $\Delta y$ as well as 
				the orientation $\theta$ between the leader and the follower.}
	\label{fig:fish_article}
\end{figure}

In \citet{Novati2017}, the authors study the swimming kinematics of two fish in 
a viscous incompressible flow. The first fish is the leader, and is affected a prescribed 
gait, while the second fish is a follower, which dynamics are unknown.
An objective of the paper is to exploit DRL to derive a swimming strategy that reduces the 
energy expenditure of the follower.
A two-dimensional Navier-Stokes equation is solved to simulate the viscous 
incompressible flow using a remeshed vortex method. To represent the fish and its 
undulation, a simplified physical model is used that describes the body curvature $k(s,t)$ 
of the fish:

\begin{equation}
\label{eq:curvature}
	k(s,t) = A(s) \sin \bigg[ 2 \pi \big( \frac{t}{T_p} - \frac{s}{L} \big) +\phi \bigg],
\end{equation}

\noindent where $L$ is the length of the fish, $T_p$ is the tail-beat
frequency, $\phi$ is a phase-difference, $t$ is the time variable, and $s$ is the abscissa. 
The observation of the environment 
provided to the DRL agent is composed of:

\begin{enumerate}
	\item The displacements $\Delta x$ and $\Delta y$ (cf figure \ref{fig:fish_article}),
	\item The orientation $\theta$ between the follower and the leader (cf figure \ref{fig:fish_article}),
	\item The moment where the action is going to take place (during one period of a tail-beat undulation),
	\item The last two actions taken by the agent.
\end{enumerate}

\noindent The agent provides actions that correspond to an additive curvature term 
for expression (\ref{eq:curvature}): 

\begin{equation*}
	k'(s,t) = A(s) M (\frac{t}{T_p} - \frac{s}{L}).
\end{equation*}

The reward to maximize is defined as $r_t = 1 - 2 \frac{\vert \Delta y \vert}{L}$, which penalizes 
the follower when it laterally strays away from the path of the leader. 
Finally, the state space is artificially restricted by terminating the current episode with 
an arbitrary reward $r_t = -1$ 
when the two fish get too far away from each other. In this application, a DQN algorithm  
is used to control the follower. For the best strategy found by the agent, swimming 
efficiency was increased by 20\% compared to baseline of multiple solitary swimmers.

\subsection{Efficient collective swimming by harnessing vortices through deep reinforcement learning \cite{Verma2018}}

\begin{figure}
	\fbox{\includegraphics[width=\linewidth]{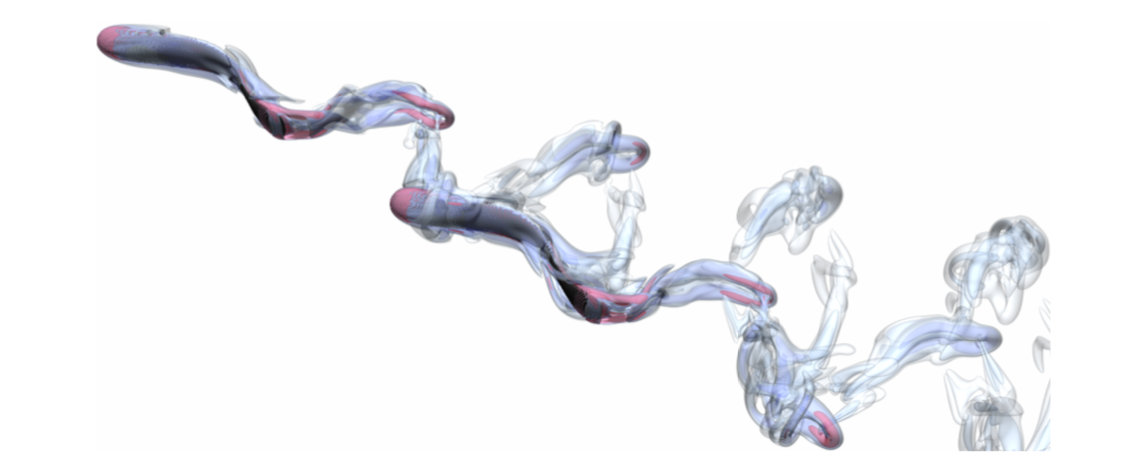}}
	\caption{\textbf{Leader and follower swimmer, reproduced from \cite{Verma2018}}}
	\label{fig:fish_article2}
\end{figure}

The authors in \cite{Verma2018} further investigated the swimming strategies of multiple fishes.
They work with two and three-dimensional flows (see figure \ref{fig:fish_article2}), and 
analyze with many details the movement of trained fishes. The DRL model is identical to 
what is used in \cite{Novati2017}, but the neural network used was improved by using 
recurrent neural networks with Long-Short Term Memory (LSTM) layers \citep{hochreiter1997}.
The authors underline the importance of this feature, arguing that here, past observations hold 
informations that are relevant for future transitions (\ie the process is not Markovian anymore).
Two different rewards are used: the first one is identical to that of last section, while the second 
one is simply equal to the swimming efficiency of the follower (see \cite{Verma2018}).
The result obtained show that collective energy-savings could be achieved with an appropriate 
use of the wake generated by other swimmers, when compared to the case of multiple 
solitary swimmers.

\subsection{Fluid directed rigid body control using deep reinforcement learning \cite{Ma2018}}

\begin{figure}
	\fbox{\includegraphics[width=\linewidth]{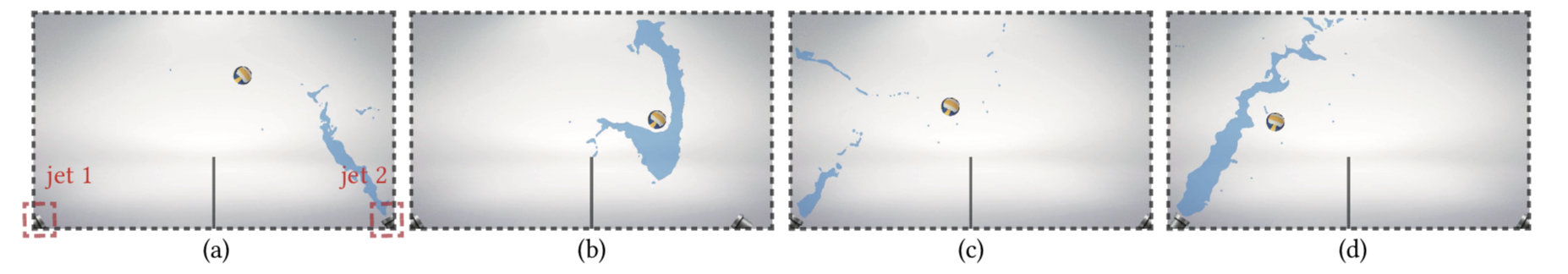}}
	\caption{\textbf{Fluid jets to control rigid body, reproduced from \cite{Ma2018}}.}
	\label{fig:fluid_solid_article}
\end{figure}

The control of complex environments that include two fluid flows interacting with multiple 
rigid bodies (cf figure \ref{fig:fluid_solid_article}) was first proposed by \cite{Ma2018}. 
In this contribution, a multiphase Navier-Stokes solver is used to simulate the full 
characteristics of the interactions between fluids and rigid bodies at different scales. 
To provide a rich set of observations to the agent while limiting its input dimension, 
a convolutional autoencoder is used to extract low-dimensional features from the 
high-dimensional velocity field of the fluid \citep{liou2014autoencoder}. This technique 
allows the use of a smaller neural network with faster convergence. The autoencoder 
is trained using a TRPO agent, which received two different observations: (i) extracted 
features from the fluid velocity field, and (ii) features from the rigid body.

In the environment, the agent is given control over several fluid jets 
(see figure \ref{fig:fluid_solid_article}), each of them described by 3 parameters: 
(i) the jet lateral acceleration $\ddot{x} _{jet}$, (ii) the jet angular acceleration 
$\ddot{\beta} _{jet}$, and (iii) the use of the jet $\delta _{jet}$. Using lateral and 
angular accelerations instead of lateral and angular positions allows the jet to 
move smoothly even with a noisy policy. Different rewards were designed 
depending on the environment goals. A possible example consists in having a rigid 
body balanced at a fixed position. In this case the designed reward is:

\begin{equation}
	r_t = w_c \exp (- || c - c^* ||^2) + w_v \exp (-||v||^2) + w_e (1 - \delta _{jet}),
\end{equation}

\noindent where the first term penalizes a position mismatch according to the goal $c^*$,
the second penalizes the presence of a residual velocity, and the last one 
encourages the agent to use the smallest amount of control forces. $w_c$, 
$w_v$ and $w_e$ are user-defined parameters used to weight the three contributions.

Overall, impressive results were obtained on complex experiments, including
playing music with the system or keeping a rotating object at a given position. 
An important feature of this contribution is the use of the autoencoder  
to extract features from the high-dimensional velocity fluid fields. Of particular interest is 
the simultaneous training of the two networks (autoencoder and policy network), 
leading to a high convergence rate of the overall method.

\subsection{Flow shape design for microfluidic devices using deep reinforcement learning \cite{Lee2018}}

\begin{figure}
	\centering
	\fbox{\includegraphics[width=\linewidth]{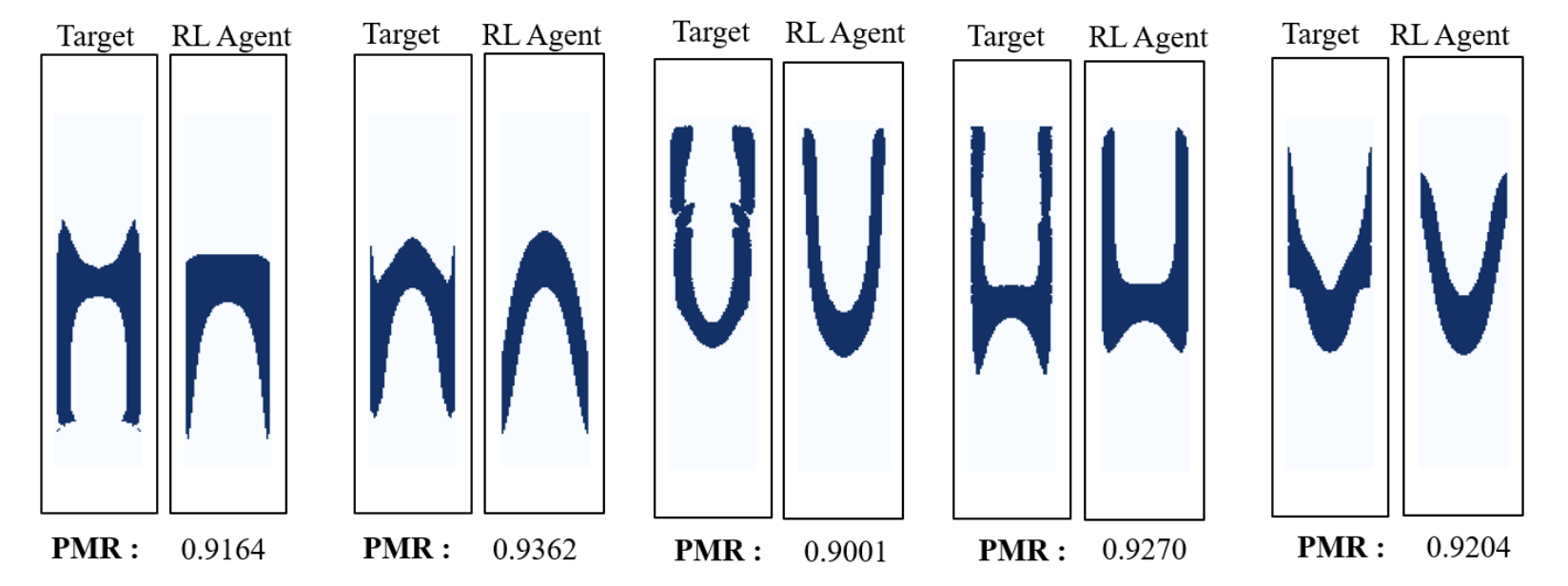}}
	\caption{\textbf{Targets flow and results obtained from DRL agent}, reproduced from \cite{Lee2018}.}
	\label{fig:pillar_flows}
\end{figure}

In \cite{Lee2018}, the authors explore the capabilities of DRL for microfluidic flow sculpting 
using pillar-shaped obstacles to deform an input flow. Given a target flow, a double DQN 
is used to generate the corresponding optimal sequence of pillars.
The observations provided to the agent consist in the current flow shape, represented 
as a matrix of black and white pixels. In return, the actions correspond to 
pillar configurations: the agent can add up to 32 pillars in order to reshape the input flow. 
A pixel match rate (PMR) function is used to evaluate the similarity between the target 
flow and the current flow. The reward is then defined using that PMR as:

\begin{equation*}
	r_t = -(1 - \frac{\text{PMR} - b}{b}),
\end{equation*}

\noindent where $b$ is the PMR of the worst-case scenario performed. PMR over 
90\% were obtained for different target flows, showing that DRL represents a viable 
alternative to other optimization processes such as genetic algorithms (see figure \ref{fig:pillar_flows}).
Transfer learning is also exploited, showing that an agent trained on a first flow could be retrained 
much more efficiently on a different flow than an untrained one, thus saving a lot of computational 
time.

\subsection{Artificial neural networks trained through deep reinforcement learning discover control strategies for active flow control \citep{Rabault2019}}

\begin{figure}
	\centering
	\fbox{\includegraphics[width=\linewidth]{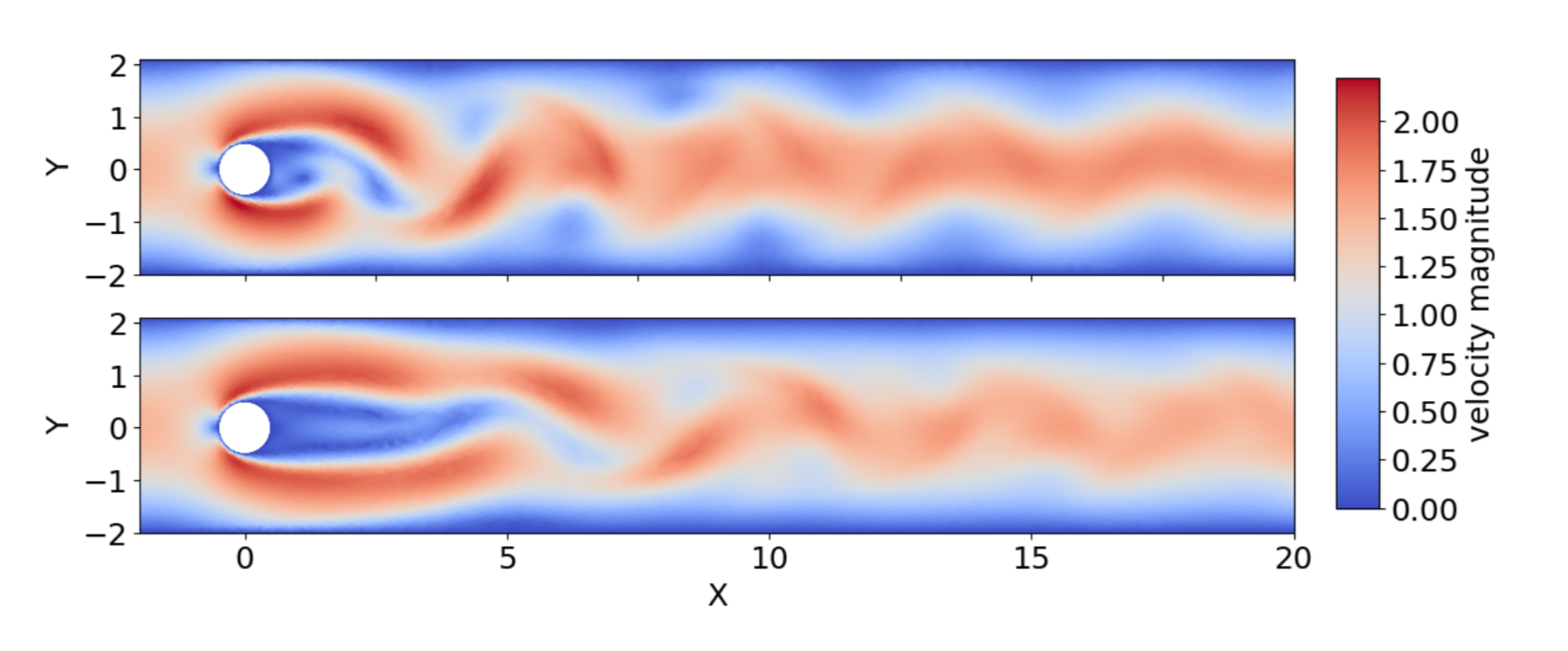}}
	\caption{\textbf{Comparison of the velocity magnitude without (top) and with
	(bottom) active flow control}, reproduced from \cite{Rabault2019}. A clear
	modification of the cylinder wake, similar to what would be obtained with
	boat-tailing, is visible.}
	\label{fig:flow_control_jet_article}
\end{figure}

In \cite{Rabault2019}, the authors present the first application of DRL for
performing active flow control in a simple CFD simulation. For this, a simple
benchmark is used. More specifically, the flow configuration is based on the
one presented in \cite{schafer1996benchmark} and features a cylinder immersed
in a 2D incompressible flow injected following a parabolic velocity profile.
The Reynolds number based on the cylinder diameter is moderate to keep the CFD
affordable, and following \citet{schafer1996benchmark}, Re=100 is used. The
incremental pressure correction scheme is used together with a finite element
approach to solve the discretized problem. A periodic Karman \textcolor{black}{vortex street} is
then obtained, as visible in figure \ref{fig:flow_control_jet_article}. Besides,
small jets are added on the sides of the cylinder in order to allow controlling
the separation of the wake. Therefore, the action performed by the DRL agent is
to decide the instantaneous mass flow rate of the small jets, while the reward
is chosen to encourage drag reduction. Finally, probes reporting the
characteristics of the flow at several fixed points in the computational domain
(either pressure or velocity can be used, both providing equivalent results)
are added in the vicinity of the cylinder and are used as the observation
provided as input to the network.

Good results are obtained, with training performed in around 1300 vortex
shedding periods. The control strategy is more complicated than traditional
harmonic forcing that had been performed in previous works
\citep{bergmann2005optimal}, which illustrates the value of using an ANN as the
controller. Typically, around 93\% of the drag induced by the vortex shedding
effect \citep{bergmann2005optimal} is suppressed by the control law found, and
the strength of the jets needed to reduce the drag is minimal. In the
established regime, the magnitude of the mass flow rate injected by the jets
normalized by the mass flow rate of the base flow intersecting the cylinder is
only of typically 0.6\%. The resulting flow, visible in figure
\ref{fig:flow_control_jet_article}, is similar to what would be
obtained with boat tailing; more specifically, an apparent increase in the
recirculation bubble is observed.

The work of \cite{Rabault2019} shows that active flow control is, at
least on a simple flow configuration, achievable with small actuations if an
adapted control law is used. While there are still many unanswered questions at
the moment, such as the robustness of the control law, the application of the
methodology to higher, more realistic (from an industrial point of view)
Reynolds numbers, or the possibility to control 3D flows, this work establishes
DRL as a new methodology that should be further investigated for the task of
performing active flow control.

\subsection{Accelerating deep reinforcement learning of active flow control strategies through a multi environment approach \citep{RabaultKuhnle2019}}
\label{section:DRL_parallel}

One of the main practical limitations of \citet{Rabault2019} lies within the
time needed to perform the learning - around 24 hours on a modern CPU. This
comes from the inherent cost of the CFD, and the fact that it is challenging to
obtain computational speedups on small tasks. In particular, increasing the
number of cores used for solving the CFD problem of \citet{Rabault2019} provides
very little speedup, as the communication between the cores negates other speed
gains. While this is particularly true on such a small computational problem,
any CFD simulation will only speed up to a maximum point independently of the
number of cores used. Therefore, a solution is needed to provide the data
necessary for the ANN to perform learning in a reasonable amount of time.

In \citet{RabaultKuhnle2019}, the collection of data used for filling the memory replay
buffer is parallelized by using an ensemble of simulations all running in
parallel, independently of each other. In practice, this means that the
knowledge obtained from several CFD simulations running in an embarrassingly
parallel fashion can be focused on the same ANN to speed up its learning.

\begin{figure}
	\centering
	\fbox{\includegraphics[width=0.49\linewidth]{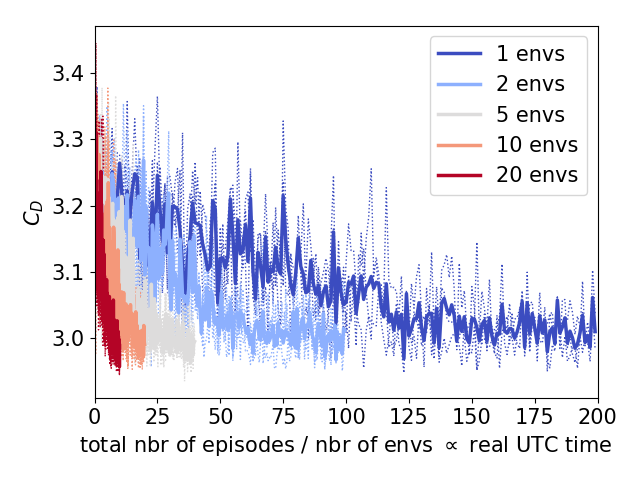}}
	\fbox{\includegraphics[width=0.49\linewidth]{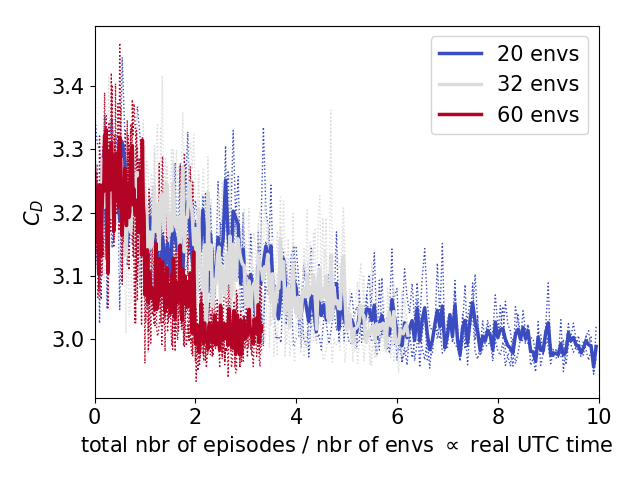}}
	\caption{\textbf{Illustration of the acceleration of ANN training through DRL using a multi
        environment approach} (reproduced from \citet{RabaultKuhnle2019}). Both plots illustrate the
        convergence of the ANN learning as a function of UTC time, depending on the number of
        environments used in parallel. In both cases, 3
    repetitions are performed for each number of environments. Thin lines indicate individual learnings. 
    The average 
    of all three learnings in each case is indicated by a thick line. In the case when the number
        of environments is a divider of the learning frequency of the ANN, the learning is 
        formally equivalent to the serial case and perfect scaling is observed (left). If a larger
        number of environments is used, some stepping is observed in the learning process, but this does
        not affect the ability to perform learning, nor the general speedup of the process (right).}
      \label{fig:accelerating_DRL_CFD_AFC}
\end{figure}

This technique results in large speedups for the learning process, as visible in figure
\ref{fig:accelerating_DRL_CFD_AFC}. In particular, for a number of
simulations that are a divider of the learning frequency of the ANN, perfect
scaling is obtained, and the training is formally equivalent to the one
obtained in serial (\ie, with one single simulation). However, even
more simulations can be used, and it is found empirically that the
resulting off-policy data sampling does not sensibly reduce the learning
ability, as visible in figure \ref{fig:accelerating_DRL_CFD_AFC}. This means
in practice that the authors achieve a training speedup factor of up to
around 60.

This natural parallelism of the learning process of ANNs trained
through DRL represents a significant advantage for this methodology, 
and a promising technical result that opens the way to the control of much more
sophisticated situations. \textcolor{black}{It is worth noting that even larger speedups can be obtained 
by making use of surrogate or reduced-order models, either pre-trained or trained on-the-fly, although specific care 
must be taken regarding the validity range of these models (as the agent could end up providing inputs to the surrogate model 
that would not match its training range).}

\subsection{Sensitivity of aerodynamic forces in laminar flows past a square cylinder}
\label{section:opt_CC}

In this section, we propose an original comparison between 
a classical optimization approach and DRL. The test case is inspired 
by \citet{Meliga2014}, where two cylinders are immersed in a moderate 
Reynolds flow. One is a square main cylinder, which position is fixed, while the other 
is a cylindrical control cylinder, which position can vary, and which radius is considered 
much smaller than that of the main cylinder. The goal of this application 
is to optimize the position of the small cylinder, such that the total drag 
of the two cylinders is inferior to that of the main cylinder alone. 
An incompressible Navier-Stokes solver \citet{Fenics2015} is used to simulate the 
flow at different Reynolds number past the square cylinder. Moreover, the flow conditions are 
identical to that described in \cite{Rabault2019}. Six different agents are trained 
in different conditions, as detailed in table \ref{tab:CC_article}.

\begin{table}
\footnotesize
\caption{\textbf{Summary of the six trained DRL agents} from \cite{garnier2019-controlcylinder}.}
\label{tab:CC_article}
\centering
\begin{tabular}{llll}
																	\toprule
\textbf{Agent} 	&  $Re$ 	& \textbf{Objective}			& \textbf{Details}			\\\midrule
1.1 			& 10 		& Control over $17$ time steps & Trained from sratch 		\\
1.2			& 10 		& Direct optimization 		& Trained from sratch 		\\\midrule
2.1 			& 40 		& Control over $17$ time steps & Transfer learning from 1.1 	\\
2.2			& 40 		& Direct optimization 		& Transfer learning from 1.2 	\\\midrule
3.1 			& 100 	& Control over $17$ time steps & Transfer learning from 1.1 	\\
3.2 			& 100 	& Direct optimization 		& Transfer learning from 1.2 	\\\bottomrule
\end{tabular}
\end{table}

\begin{figure}
	\centering
	\includegraphics[width=.75\linewidth]{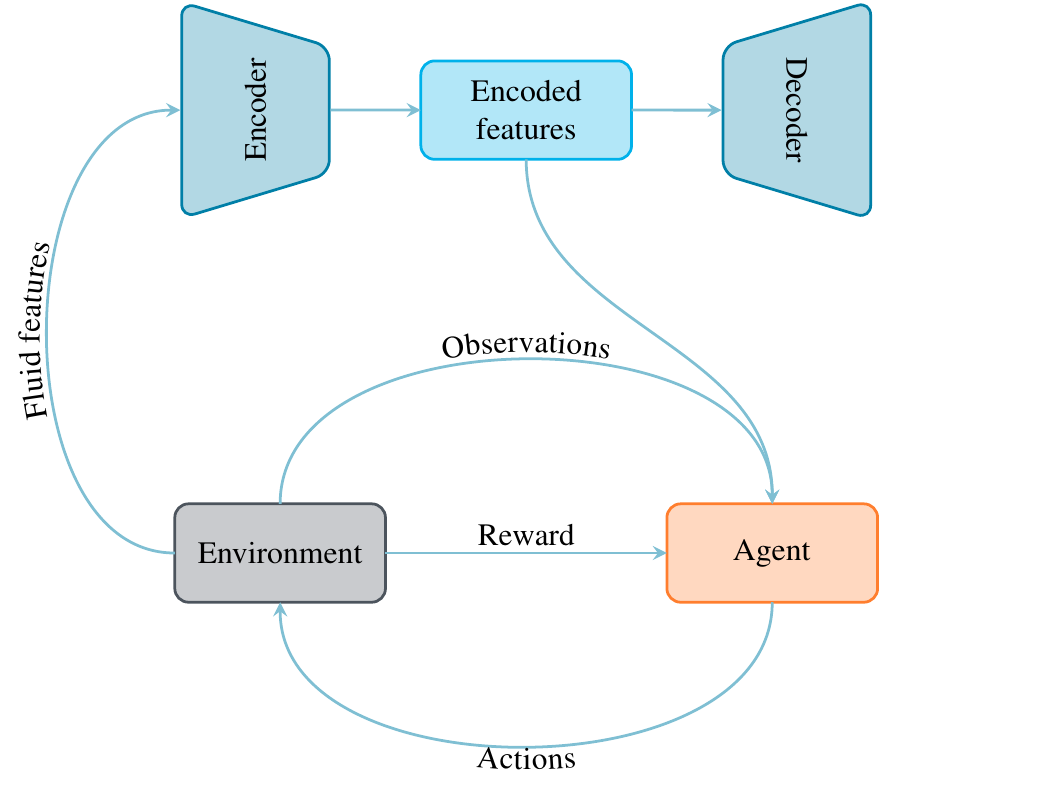}
	\caption{\textbf{Architecture for agents 1.1, 2.1 and 3.1}.}
	\label{fig:CC_architecture2}
\end{figure}

Two different architectures are proposed. The first one is a classical DRL architecture, 
similar to what is shown in figure \ref{fig:summary_rl}, but with a difference that episodes were 
reduced to a single action. In that case, the agent directly attempts to propose an 
optimal $(x,y)$ position from the same initial state, using a PPO algorithm. 
On the contrary, in the second architecture, successive displacements $(\Delta x, \Delta y)$ 
of the control cylinder are 
provided by an A3C agent, starting from a random initial position. The observations provided to the 
agent are made of the last two positions 
of the control cylinder, the Reynolds number and encoded features from the velocity 
fields. To that end, an autoencoder was trained simultaneously to the agent to reduce the 
state dimensionality (with a dimension reduction from around 8000 to 70), in a similar fashion to \cite{Ma2018} 
(see figure \ref{fig:CC_architecture2}). The same reward function is used for both architectures: 

\begin{equation*}
	r_t = C_{D}^{\,0} - C_D,
\end{equation*}

\noindent where $C_{D}^{\,0}$ is the drag of the main cylinder alone, $C_{D}$ is the total drag of the main 
and the control cylinders, and the drag coefficients are counted positively. With this simple expression, 
a positive reward is obtained when the combined drag of the cylinders is inferior to the 
reference drag value.

\begin{figure}
	\centering
	\begin{subfigure}[t]{.46\textwidth}
		\centering
		\fbox{\includegraphics[height=.15\textheight]{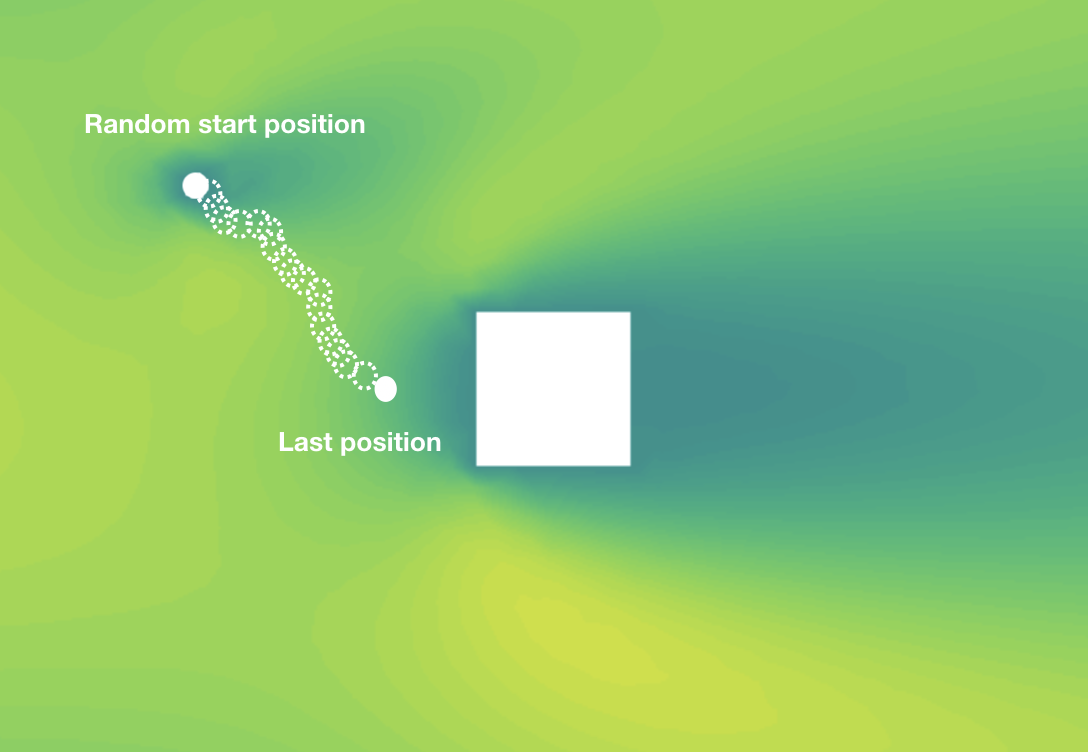}}
		\caption{Small displacements, $Re = 10$ (agent 1.1)}
	\end{subfigure}
	\qquad
	\begin{subfigure}[t]{.45\textwidth}
		\centering
		\fbox{\includegraphics[height=.15\textheight]{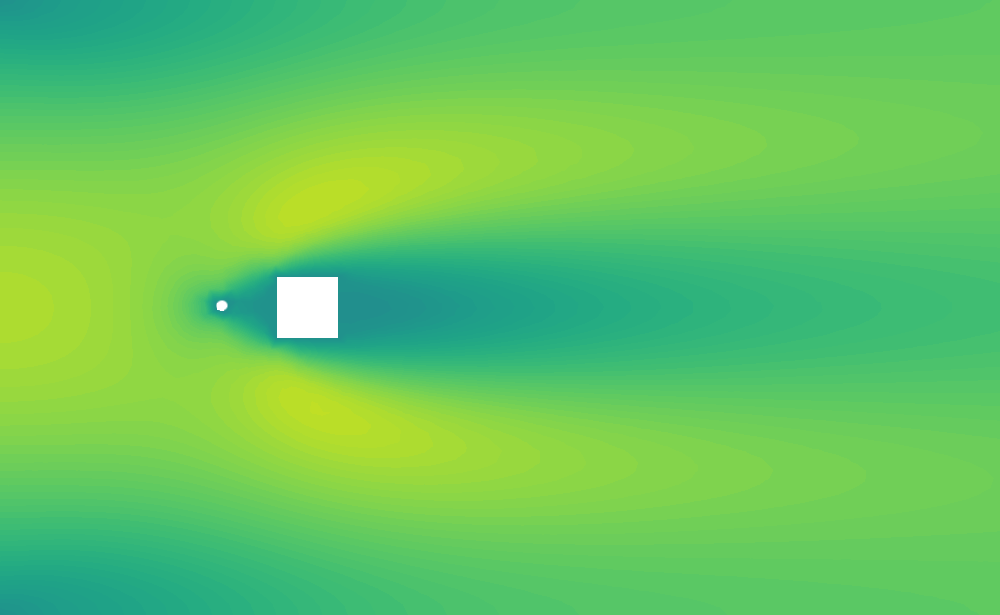}}
		\caption{Direct optimization, $Re = 10$ (agent 1.2)}
	\end{subfigure}
	
	\bigskip
	
	\begin{subfigure}[t]{.46\textwidth}
		\centering
		\fbox{\includegraphics[height=.15\textheight]{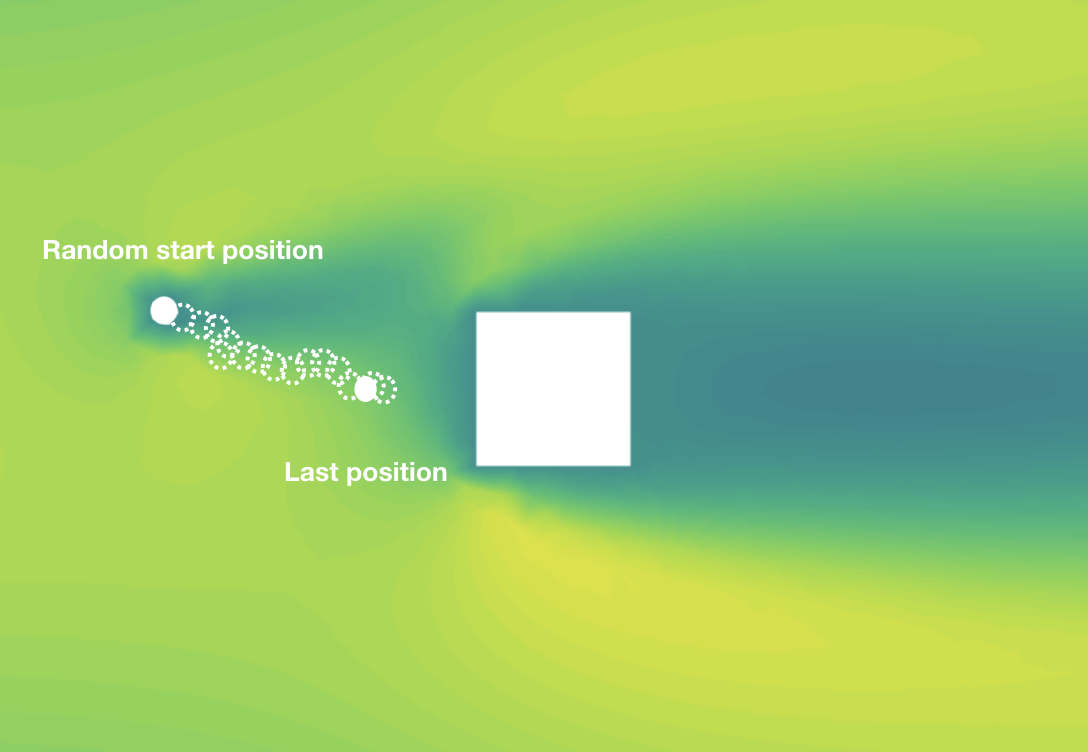}}
		\caption{Small displacements, $Re = 40$ (agent 2.1)}
	\end{subfigure}
	\qquad
	\begin{subfigure}[t]{.45\textwidth}
		\centering
		\fbox{\includegraphics[height=.15\textheight]{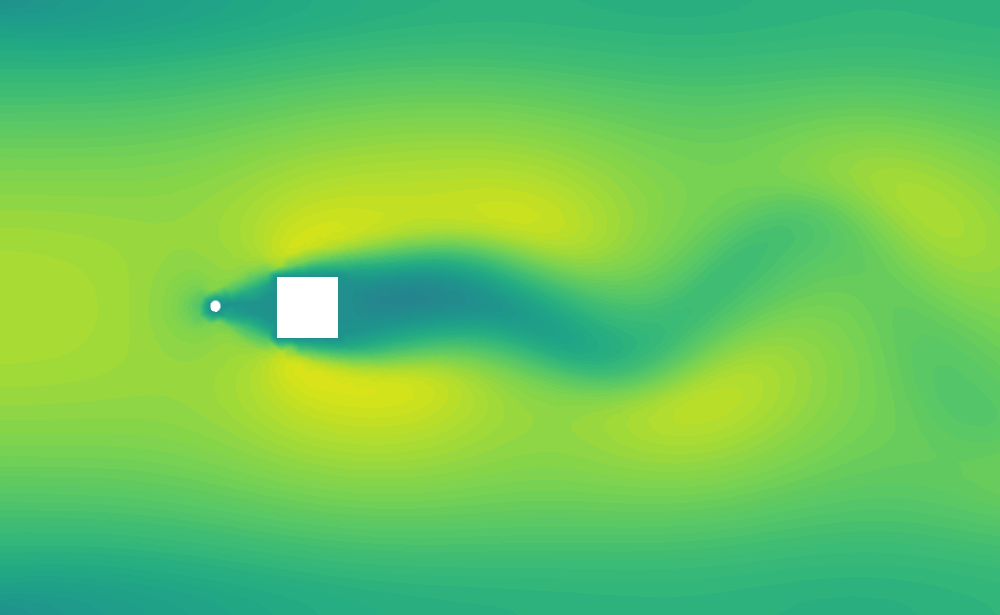}}
		\caption{Direct optimization, $Re = 40$ (agent 2.2)}
	\end{subfigure}
	
	\bigskip
	
	\begin{subfigure}[t]{.46\textwidth}
		\centering
		\fbox{\includegraphics[height=.15\textheight]{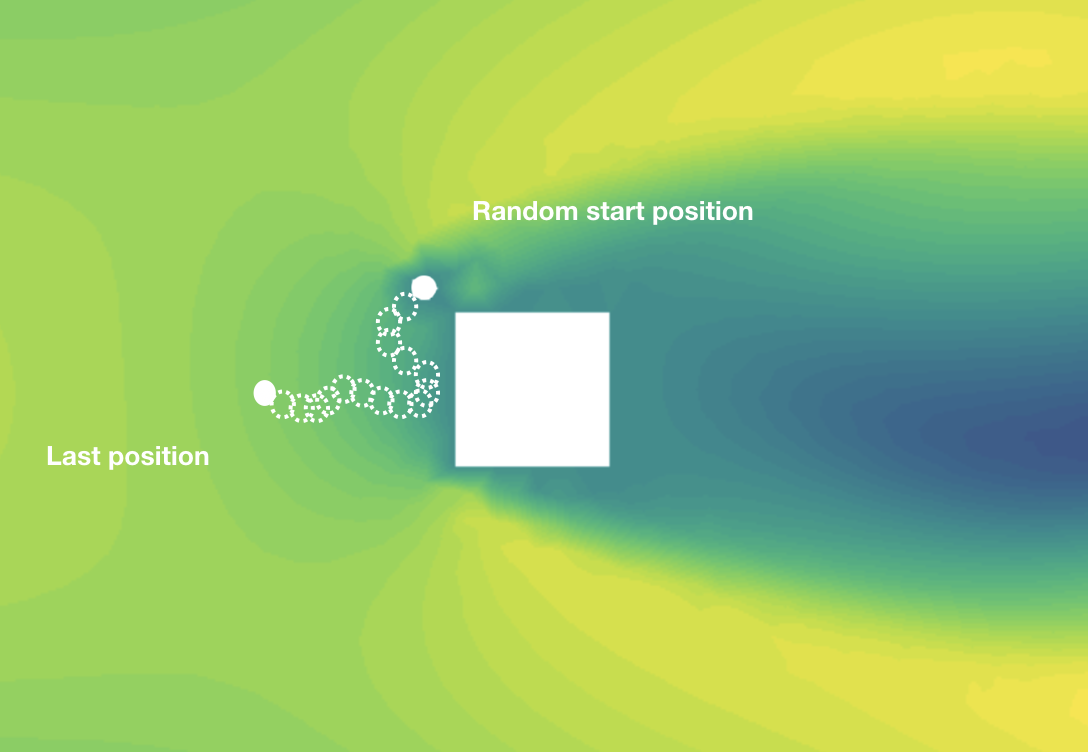}}
		\caption{Small displacements, $Re = 100$ (agent 3.1)}
	\end{subfigure}
	\qquad
	\begin{subfigure}[t]{.45\textwidth}
		\centering
		\fbox{\includegraphics[height=.15\textheight]{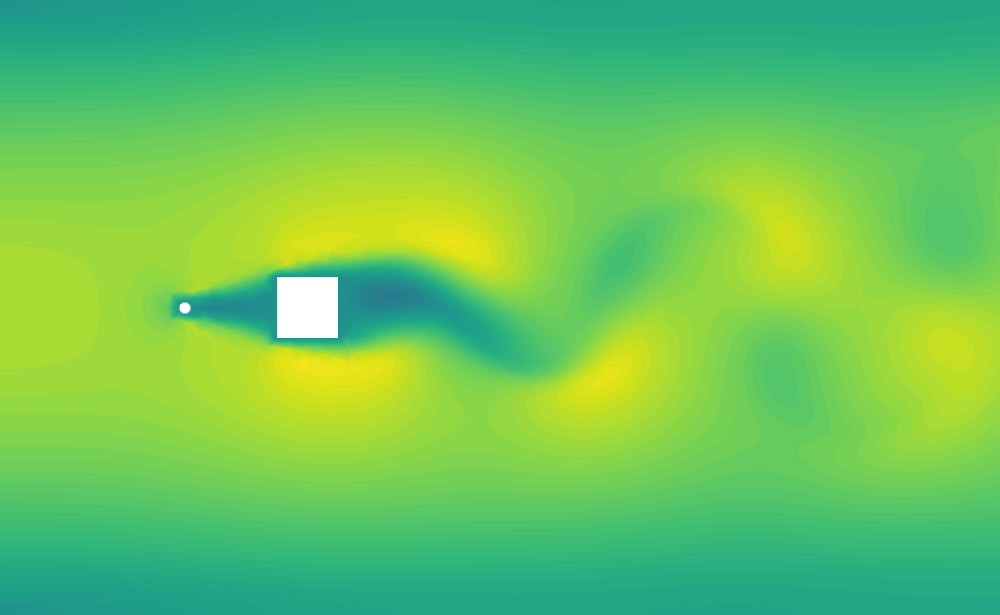}}
		\caption{Direct optimization, $Re = 100$ (agent 3.2)}
	\end{subfigure}
	\caption{\textbf{Results obtained by the trained agents}. The environment consists of a square of unit lateral size centered in $(0,0)$, immersed in a rectangular domain of dimension $\left[ -5, 20\right] \times \left[ -5, 5\right]$.}
	\label{fig:CC_direct_results}
\end{figure}

We compare the results obtained with those 
from \cite{Meliga2014}, where a classical adjoint method is used. Overall, the same 
optimal positions were found for the control cylinder, at $Re = 40$ and $Re = 100$.
Although the case $Re = 10$ was not present in the original paper, coherent results 
were obtained for the configuration too. For the first agents at $Re = 10$,
training took approximately $5$ hours using parallel learning with 17 CPUs. 
Regarding the last four agents, the transfer learning allowed them to
learn at a much faster pace: only $10$ hours were necessary on a single CPU. An 
overview of the results is shown in figure \ref{fig:CC_direct_results}.

With this experiment, we showed that not only DRL could perform both control and
direct optimization tasks, but also that transfer learning between different
configurations represents a powerful feature to save computational time. We also confirmed that 
the use of an autoencoder is a safe and robust way to extract essential features
from complex and high dimensional fluid fields. Details about this test case and 
the code used to reproduce these results can be found 
on the GitHub repository: \url{https://github.com/DonsetPG/fenics-DRL}. 
We recall that this code uses different library:

\begin{enumerate}
	\item FEniCS \citep{Fenics2015} for the \textcolor{black}{CFD} solver,
	\item Gym \citep{Gym2016} for the DRL environment,
	\item Stable-baselines \citep{stable-baselines} for the DRL algorithms.
\end{enumerate}

Future work on this library will focus on generalizing the coupling of DRL with different open source CFD code.

\section{Conclusion}

Although DRL has already been applied to several cases of optimization and 
control in the context of fluid dynamics, the literature on the topic remains particularly shallow. 
In the present article, we reviewed the available contributions on the topic to provide 
the reader with a comprehensive state of play of the possibilities of DRL in fluid mechanics.
In each of them, details were provided on the numerical context and the problem complexity.
The choices of the DRL algorithm and the reward shaping were also described.

Given the high-level interfaces of existing libraries, the coupling of DRL algorithms with 
existing numerical CFD solvers can be achieved with a minimal investment, while opening 
a wide range of possibilities in terms of optimization and control tasks. The algorithms 
presented in this review proved to be robust when exposed to possible numerical noise, 
although high $Re$ applications remain to be achieved. Additionally, the parallel 
capabilities of the main DRL libraries represent a major asset in the context of time-expensive 
CFD computations. Transfer learning also proved to be a key feature in saving 
computational time, as re-training agents already exploited in similar situations led to 
a fast convergence of the agent policy. The use of autoencoders to feed the agent with both a 
compact and rich observation of the environment also showed to be beneficial, as it allows for 
a reduced size of the agent network, thus implying a faster convergence.

As of now, the capabilities and robustness of DRL algorithms in highly turbulent and non-linear flows 
remain to be explored. Also, their behavior and convergence speed in action spaces of 
high dimensionalities is unknown. It makes no doubt that the upcoming years will see the mastering of these 
obstacles, supported by the constant progress made in the DRL field and driven by 
the numerous industrial challenges that could benefit from it.

\bibliography{biblio} 
\bibliographystyle{jfm}

\end{document}